%% file: tully.tex
\documentclass[graybox]{svmult}

\usepackage{mathptmx}       % selects Times Roman as basic font
\usepackage{helvet}         % selects Helvetica as sans-serif font
\usepackage{courier}        % selects Courier as typewriter font
\usepackage{type1cm}        % activate if the above 3 fonts are
\usepackage{makeidx}         % allows index generation
\usepackage{graphicx}        % standard LaTeX graphics tool
\usepackage{multicol}        % used for the two-column index
\usepackage[bottom]{footmisc}% places footnotes at page bottom

\usepackage{natbib}

\makeindex             % used for the subject index
\begin{document}
\input{defs}
\title*{Lighthouses in the Shoals of Dark Halos.  
Conference Honoring Ken Freeman's 70th Birthday: `Galaxies and their Masks', Sossusvlei, Namibia 12-16 April, 2010}
\titlerunning{Galaxies: Lighthouses in the Shoals of Dark Halos} 
\author{R. Brent Tully}
\institute{R. B. Tully \at Institute for Astronomy, University of Hawaii, Honolulu,Hawaii, \email{tully@ifa.hawaii.edu}}
\maketitle

\abstract{It is anticipated from hierarchical clustering theory that there are scaling relationships between halos over a wide range of mass.  Observationally it can be difficult to identify the markers that characterize these relationships because of the small numbers of visible probes and confusion from contaminants in projection.  Nonetheless, in favorable circumstances it is possible to identify a very useful marker: the radius of the caustic at second turnaround.  In a few favorable circumstances it is possible to identify the radius of first turnaround, or zero velocity surface about a collapsed region.  It will be shown that specifically the radius of second turnaround scales as anticipated over three orders of magnitude in mass from $10^{12}$ to $10^{15}~M_{\odot}$.  Halos are characterized by zones of dispersed velocities within the second turnaround caustic and zones of infall between the first and second turnaround radii.  The inner zone is populated in the majority by gas poor morphologies and the outer zone is populated in the majority by gas rich morphologies.  The numbers of dwarfs within the inner zone is roughly constant per unit halo mass.}

\section{Introduction}
\label{sec:1}

In the imaginary world of simulations, researchers have a well developed picture of the collapse of matter into halos.  Over time, small halos are absorbed into larger units.  Collapsed regions filled with substructure can be defined by hundreds or thousands of particles.  Halos can be identified with precision within those simulations \citep{2009MNRAS.398.1150B}.

In the real world, most galaxies are observed to lie in groups or clusters.  However, membership may be so limited that the structure is ill defined.  There is no concensus among observers about what is meant by the terms `group' and `cluster'.  For example, it is commonly accepted that we live in something called the Local Group.  A quantitative boundary that would include the traditional members is the zero--velocity surface  \citep{2009MNRAS.393.1265K}.  Galaxies inside this surface are infalling or on more complex bound orbits.  Galaxies outside this surface are participating in the Hubble expansion.  But now consider the Virgo Cluster.  This entity would traditionally be taken to include the region within 2 Mpc where several thousand galaxies are following randomized orbits \citep{1987AJ.....94..251B}. The zero--velocity surface around the Virgo Cluster lies at a radius of $\sim 7$ Mpc, almost half of our distance of 16.7 Mpc from the cluster.  The implicit traditional definitions of the Local Group and the Virgo Cluster are inconsistent.

It does not seem useful to draw a distinction between the terms `cluster' and `group'.  Clusters contain a lot of galaxies and groups contain only a few, but there is no clear demarcation between the two.  In this article, the two terms will be used interchangeably.  

So how should an observer define a group?   Structure forms as sufficient matter accumulates through gravitational attraction to decouple from the expansion of the universe.  The matter falls together on nearly radial orbits and, given enough time, evolves toward dynamic equilibrium.  Two natural dimensions to describe a collapsed structure are the gravitational or virial radius, $r_g$ and the radius at 200 times the `critical' density for a matter-dominated closed universe, $r_{200}$.  Here, lower case $r$ implies a 3-dimensional radius and an upper case $R$ will be used to denote a projected radius.  Accordingly, if all galaxies in a group sample are given equal weight (ie, galaxies of whatever luminosity are considered to be test particles within an environment dominated by distributed dark matter) then the virial radius is defined as 
\begin{equation}
R_g = {{\rm N}^2 \over \sum_{i<j} 1 / R_{ij}}
\label{rg}
\end{equation}
where $R_{ij}$ is the projected distance between pairs, and a virial mass estimate is given for the group by
\begin{equation}
M_{\rm v} = \alpha \sigma_p^2 r_g / G = (\pi/2) \alpha \sigma_p^2 R_g / G
\end{equation}
where the line-of-sight velocity dispersion for the sample of N galaxies is
\begin{equation}
\sigma_p = \sqrt{\sum_i (v_i - <v>)^2 / {\rm N}} ~.
\end{equation}
The group mean velocity is $<v>$ and $\alpha = 3$ if orbits are isotropically distributed.

The alternative dimension $r_{200}$ is defined to coincide with a density $200 \rho_{crit}$ where $\rho_{crit} = 3 {\rm H}_0^2 / 8 \pi G$ at redshift zero.  This parameter can be determined in simulations with large numbers of test particles but it is not such a useful construct in the context of observations of small groups.  However,  \citet{1999ApJ...518...69M} calculate a ratio between $r_{200}$ and $r_g$ assuming $M(r) \propto r$ that can be reduced to the relationship (at $z=0$)
\begin{equation}
r_{200} = {\sqrt{\alpha} \sigma_p \over 10 {\rm H}_0} ~.
\end{equation} 

There may be other dimensions that are observationally useful.
Consider the collapse of a spherically symmetric overdense region in the expanding universe  \citep{1972ApJ...176....1G,1985ApJS...58...39B}.  The time of collapse, $t_c$, depends on density, $\rho$: $t_c \propto \rho^{-1/2}$.  In the approximation of spherical collapse, all across the universe at
$t_c = today$ overdense regions with a common density will be separating from the cosmic expansion, creating zero--velocity surfaces.  The radius that encloses one of these regions will be called the first turnaround radius, $r_{1t}$.

At earlier times, regions of successively higher density collapsed.  There was a time and a corresponding density that lead to collapse and re-expansion to a pause before recollapse.  This pause and turnaround creates a caustic  \citep{1989RvMP...61..185S} at what will be called the second turnaround radius, $r_{2t}$.

At yet earlier times, regions of yet higher density collapsed and resulted in caustics of higher order turnarounds.  In the real world, departures from spherical symmetry would result in violent relaxation  \citep{1967MNRAS.136..101L}, departures from radial orbits, and blurring of the caustics.  The collapsed region will begin to evolve toward energy equipartition.  Such regions will be referred to as `quasi-virialized'. 

These theoretical musings provide a basis for a definition of groups.  It will be argued that radii of second turnaround and, more problematically of first turnaround, are observable.

The caustic of second turnaround for a group can potentially be defined by two features: a density discontinuity and a velocity dispersion discontinuity.  These features were displayed by  \citet{1985ApJS...58...39B} in the simple case of spherical infall of an initial top hat density excess.  In this simple case, density falls off as a power law with radius within $r_{2t}$ and velocity dispersions are large.  Beyond $r_{2t}$, the density is expected to take a downward step and velocities locally are expected to manifest coherent infall with modest dispersion.  The same salient features are recovered with N-body simulations based on the same spherical infall model \citep{2009MNRAS.400.2174V}.
These features might be hard to identify in observational information.  The density step might be obscured by noise from line-of-sight contamination.  The step from large to small velocity dispersions can easily be confused by the limitation of access to only radial velocities and lack of precise distances that would distinguish backside from frontside infall.  The situation with many groups will be messy but there are relatively clean cases where the desired information can be recovered.  For example one can identify the caustics in redshift--radius plots for rich clusters \citep{1997ApJ...481..633D, 2003AJ....126.2152R}.  The challenge given attention here is to find analogous features in groups with much smaller populations.  The challenge can be met if the entities are sufficiently isolated or so nearby that their three dimensional structure can be discerned. 

The musings suggest that there should be interesting scaling relations between groups.  In the spherical approximation, the density at $r_{2t}$ is the same for all groups
\begin{equation}
{M(r_{2t}) \over r_{2t}^3} = constant.
\end{equation} 
With the approximation $M(r) \propto r$ 
\begin{equation}
M(r_{2t}) = M_{\rm v} (r_{2t} / r_g)
\end{equation} 
so
\begin{equation}
{M_{\rm v} \over r_{2t}^3} \cdot {r_{2t} \over r_g} = {\sigma_p^2 \over r_{2t}^2} = constant.
\end{equation} 
Hence
\begin{equation}
r_{2t} \propto \sigma_p \propto M_{\rm v}^{1/3}.
\end{equation}
In subsequent sections, data will be presented that confirm these relationships and provide correlation zero points.

There is the prospect that equivalent scaling relationships can be found based on the first turnaround radius, $r_{1t}$.  The interest here is that the ratio $r_{1t}/r_{2t}$ is dependent on the dark energy content of the universe.  Collapse conditions at a given time depend almost exclusively on the matter density but dark energy affects the clock, hence the timing between collapse events.  In a flat universe with 70\% of the cosmological density in vacuum energy then $r_{1t}/r_{2t} = 3.3$ whereas in a flat universe with the critical density in matter then $r_{1t}/r_{2t} = 3.7$ (Gary Mamon, private communication). 

Finally it is to be entertained if there are observational manifestations of the `zero gravity' surface at a radius $r_{zg}$ around a halo.  Only material within this surface will ever participate in collapse onto the halo.  It has been suggested \citep{2009A&A...507.1271C} that the region of the zero gravity surface is expected to be evacuated and can be identified by this property.  The importance of this feature depends on the properties of dark energy.

\section{From Big to Small}

\begin{figure}[t]
\sidecaption
% For example, with the graphicx style use
\includegraphics[scale=.3]{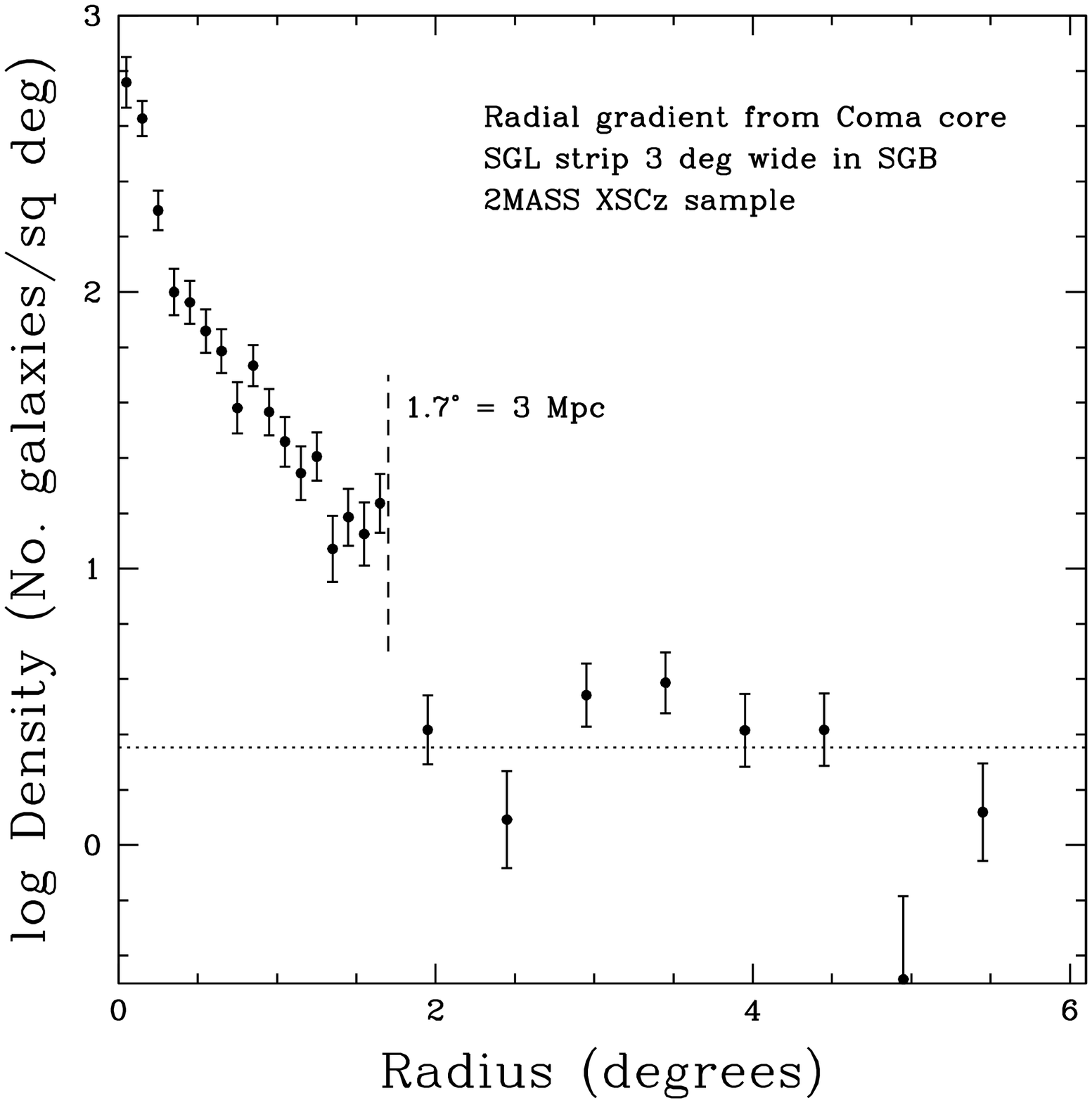}
\includegraphics[scale=.3]{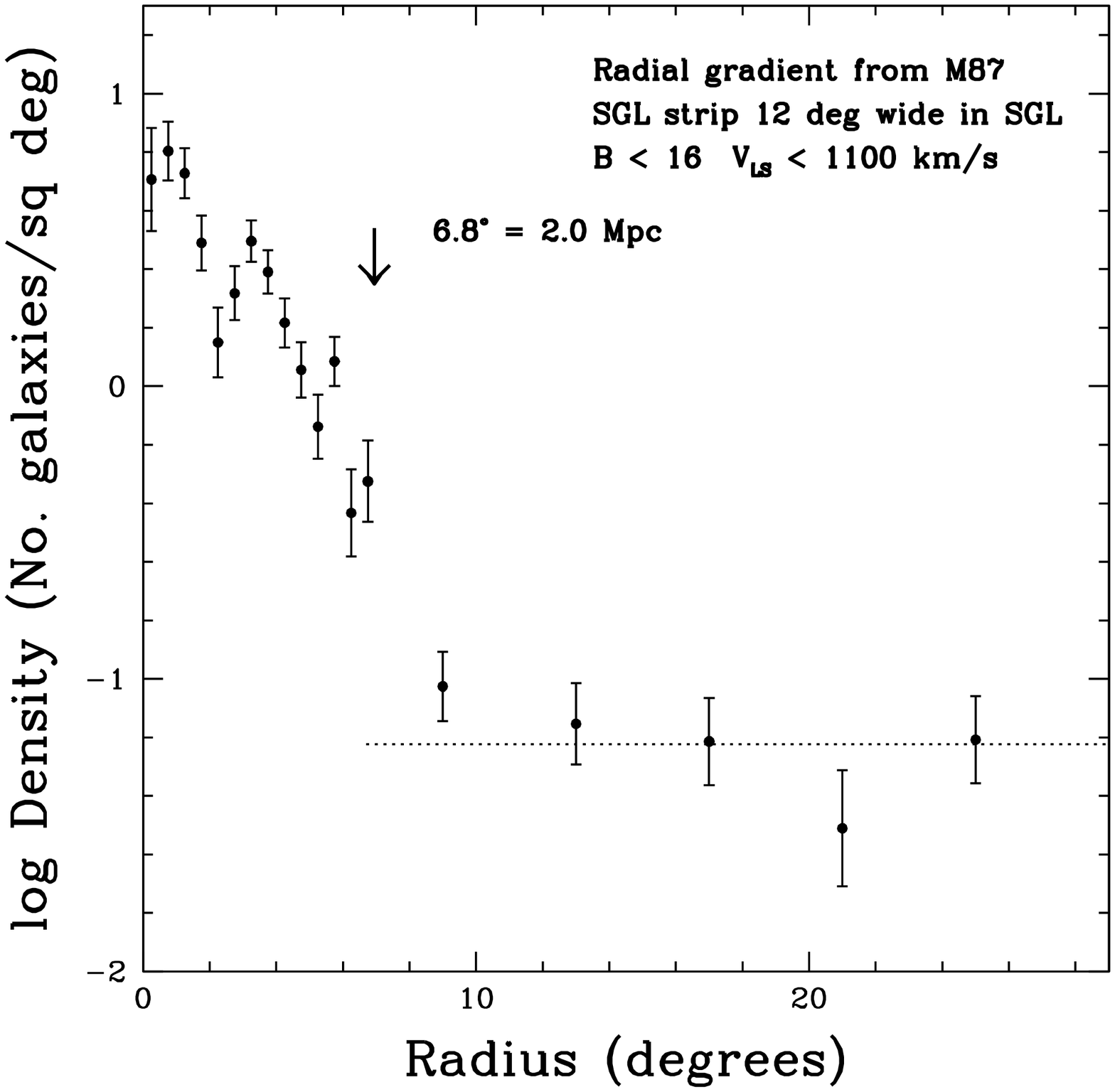}
%
% If no graphics program available, insert a blank space i.e. use
%\picplace{5cm}{2cm} % Give the correct figure height and width in cm
%
\caption{Radial distributions of galaxies in the Coma and Virgo clusters.  The density discontinuities at 3 Mpc and 2.0 Mpc respectively are associated with the caustics of second turnaround.}
\label{coma-virgo}       % Give a unique label
\end{figure}

The questions to be addressed are whether there are observational features that distinguish the quasi-virialized and infall regions of halos and, if so, how these features scale with halo mass.  Most of the discussion in this paper will focus on modest to very small groups of galaxies, the visible manifestations for the overwhelming majority of identifiable halos.   The sites of intermediate mass halos, those in the range $5 \cdot 10^{12} - 10^{14} ~\Msun$, have been investigated with a wide field imaging survey at the Canada-France-Hawaii Telescope (CFHT), initially using the 0.3 sq. deg. 12K CCD camera and then the 1 sq. deg. Megacam detector.  \citep{2009AJ....138..332J}.  Follow up spectroscopy was undertaken with Subaru and Keck telescopes.   The study is extended to very small halos in the range $10^{11} - 5 \cdot 10^{12}~\Msun$ by giving consideration to the region within 4~Mpc where, outside the zone of obscuration, almost every galaxy brighter than $M_B = -11$ has probably been identified \citep{2004AJ....127.2031K}.  Almost all of these nearby galaxies have now been sufficiently observed with Hubble Space Telescope (HST) that an accurate distance is available from a measurement of the luminosity of the tip of the red giant branch, the TRGB method     There is now detailed information on the grouping properties of galaxies down to the scales of associations of dwarfs \citep{2006AJ....132..729T}.    Page limitations prevent a detailed discussion of the many individual group environments that have been given attention.   The following is an expurgated version of a study that will be flushed out in a journal article. 

The Coma Cluster rests in the nearest massive and evolved halo at $2 \cdot 10^{15}~\Msun$ and the Virgo Cluster is the dominant structure in the Local Supercluster at $8 \cdot 10^{14}~\Msun$.  Each is continuing to grow by infall along filaments.  The dimensions orthogonal to these filaments are well defined.  It is shown in Figure~\ref{coma-virgo} that density discontinuities give good definition of the caustics of second turnaround.  In the case of the Virgo Cluster where there is improving distance resolution the location of the caustic is supported by velocity information.

\begin{figure}[]
\sidecaption
\includegraphics[scale=.3]{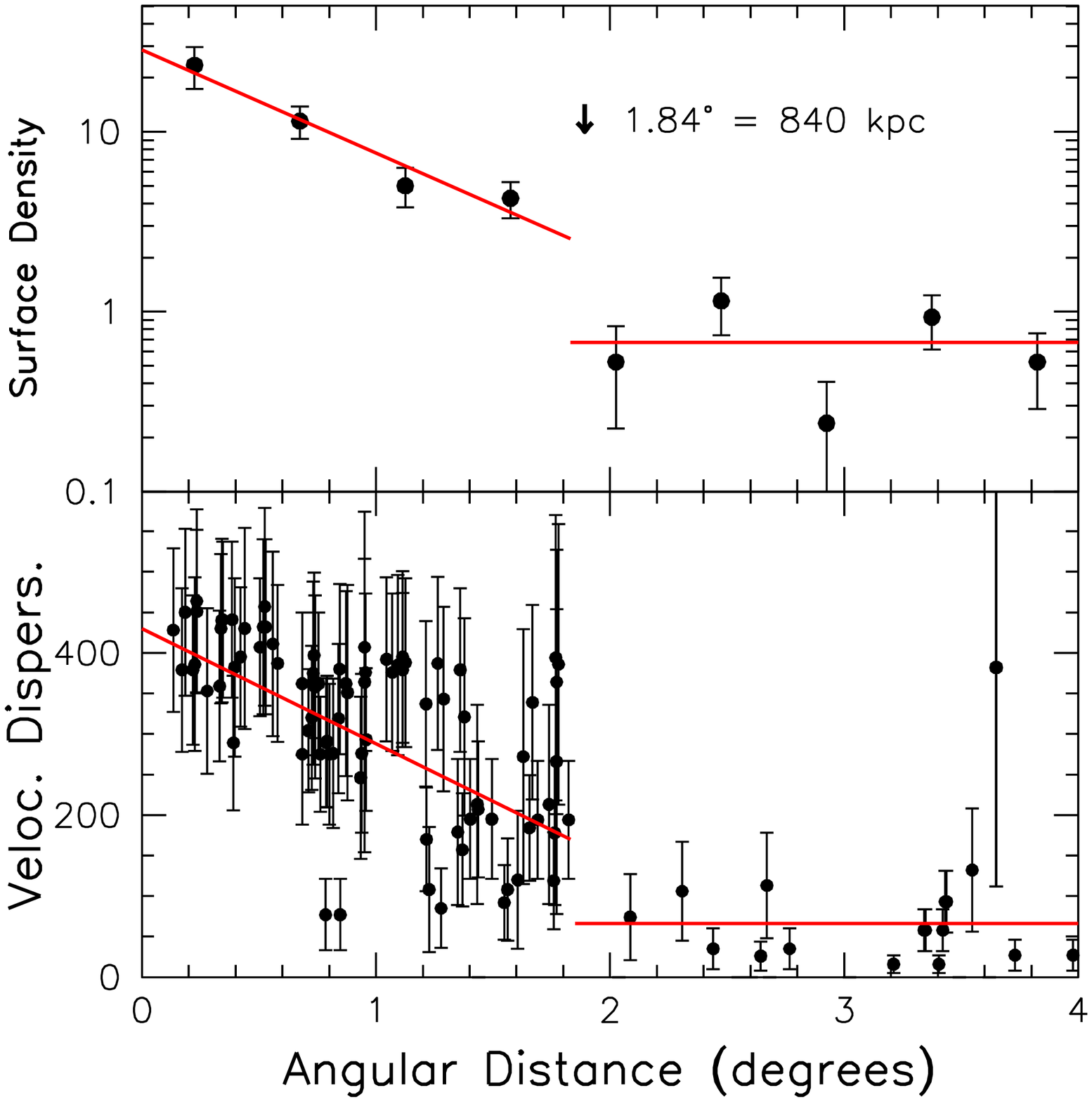}
\includegraphics[scale=.31]{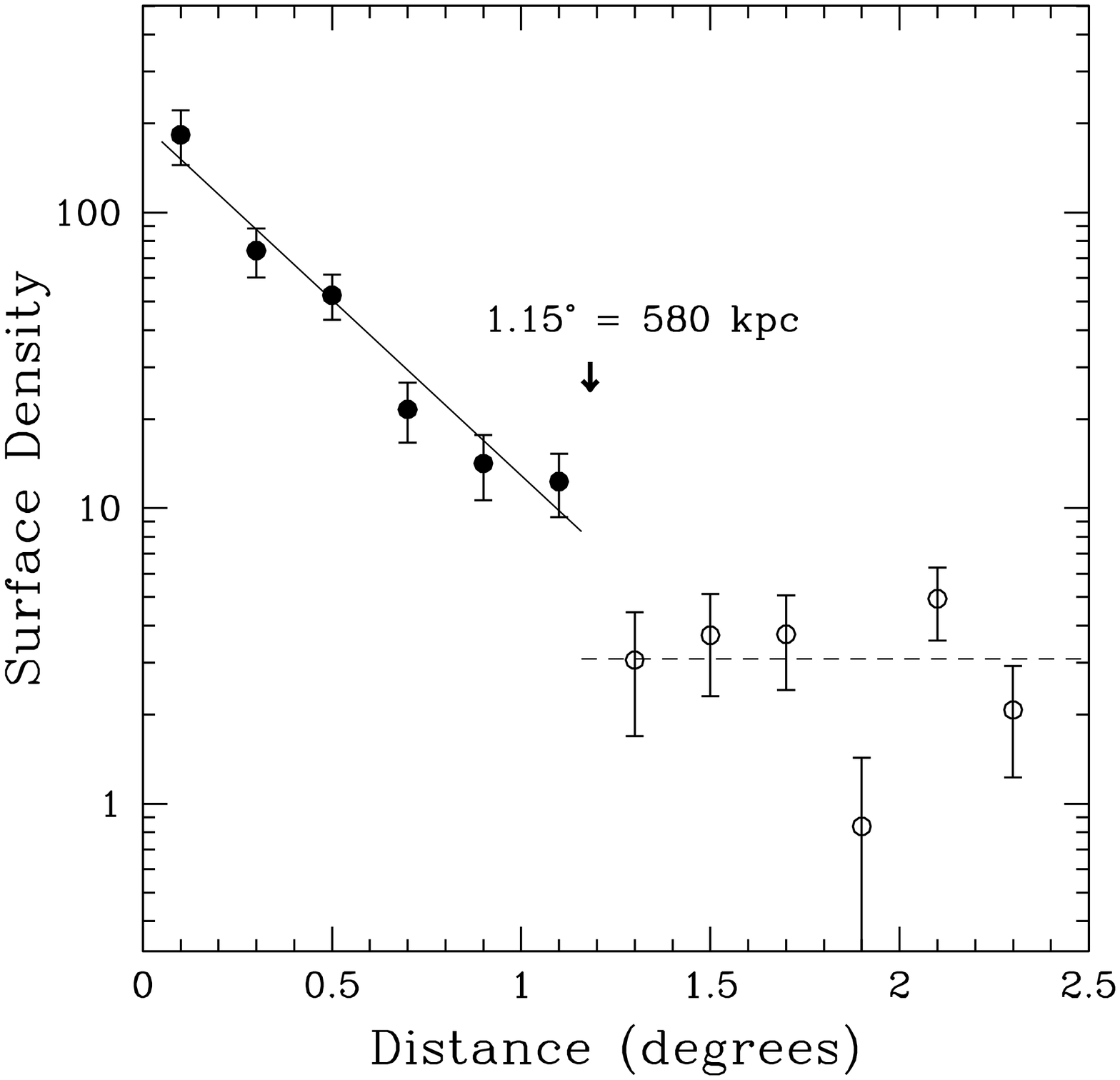}
\caption{Radial distributions of galaxies in the NGC 5846 and N5353/4 groups.  The density discontinuities at 840 kpc and 580 kpc respectively are associated with the caustics of second turnaround. The velocity discontinuity at the caustic is shown in the case of NGC 5846.}
\label{n5846-n5371} 
\end{figure}

The CFHT wide field imaging survey gave attention to three groups with masses in the range $2-8 \cdot 10^{13}~\Msun$ at distances $25-29$~Mpc \citep{2005AJ....130.1502M, 2006MNRAS.369.1375T, 2008AJ....135.1488T}.  The evidence for the detection of second turnaround caustics is shown for two of these cases in Figure~\ref{n5846-n5371}.  Two of these intermediate mass halos are dynamically evolved, with central dominant ellipticals and gas-poor companions, while the third presents a situation with mixed morphologies.

\begin{figure}[]
\sidecaption
\includegraphics[scale=.38]{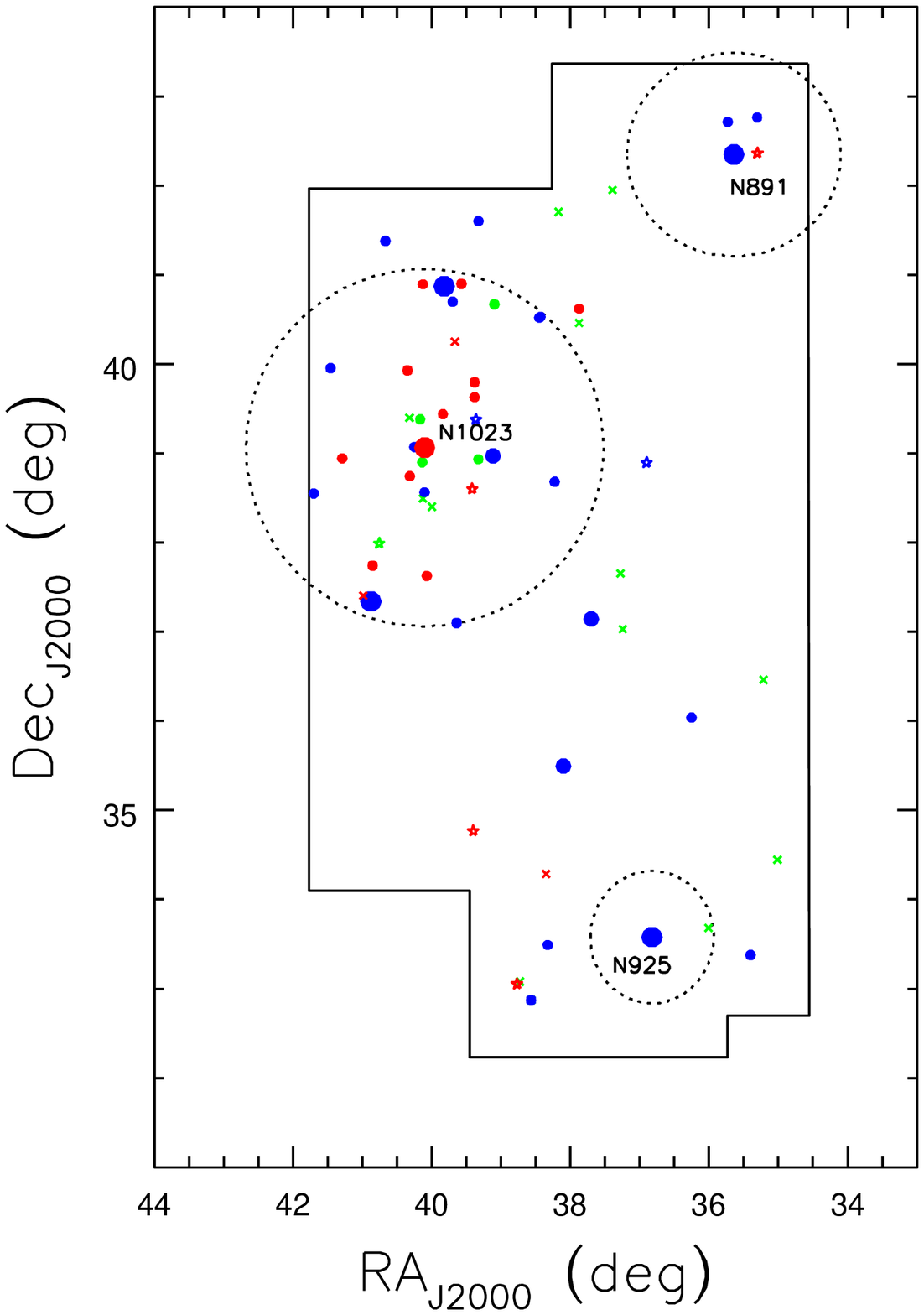}
\includegraphics[scale=.3]{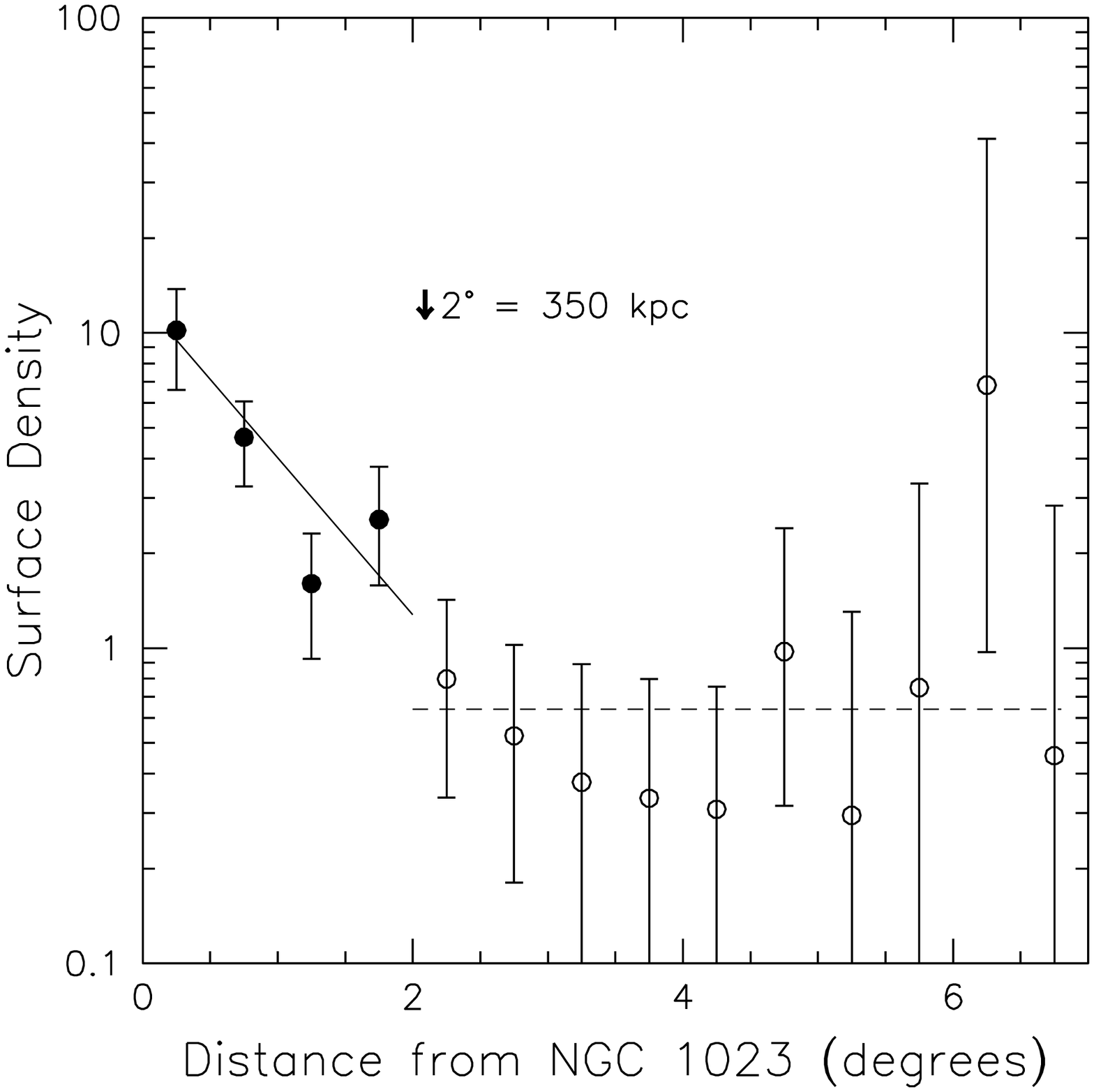}
\caption{{\it Left:} Galaxies in the vicinity of NGC 1023. Red: early; blue: late; green: transition.  Irregular box: region of CFHT wide field imaging survey.  Dotted circles: inferred caustics of second turnaround for the 3 dominant halos.  {\it Right:} Radial distribution of galaxies with respect to NGC 1023.  The caustic of second turnaround is associated with the discontinuity at 350 kpc.}
\label{n1023} 
\end{figure}

In an effort to understand what is going on in low density environments, a substantial area around the galaxy NGC~1023 was imaged with the CFHT cameras \citep{2009MNRAS.398..722T}.  The central halo has a mass of $6 \cdot 10^{12}~\Msun$.  The detection of the second turnaround caustic and the morphological segregation of galaxies is seen in Figure~\ref{n1023}.

\begin{figure}[]
\sidecaption
\includegraphics[scale=.3]{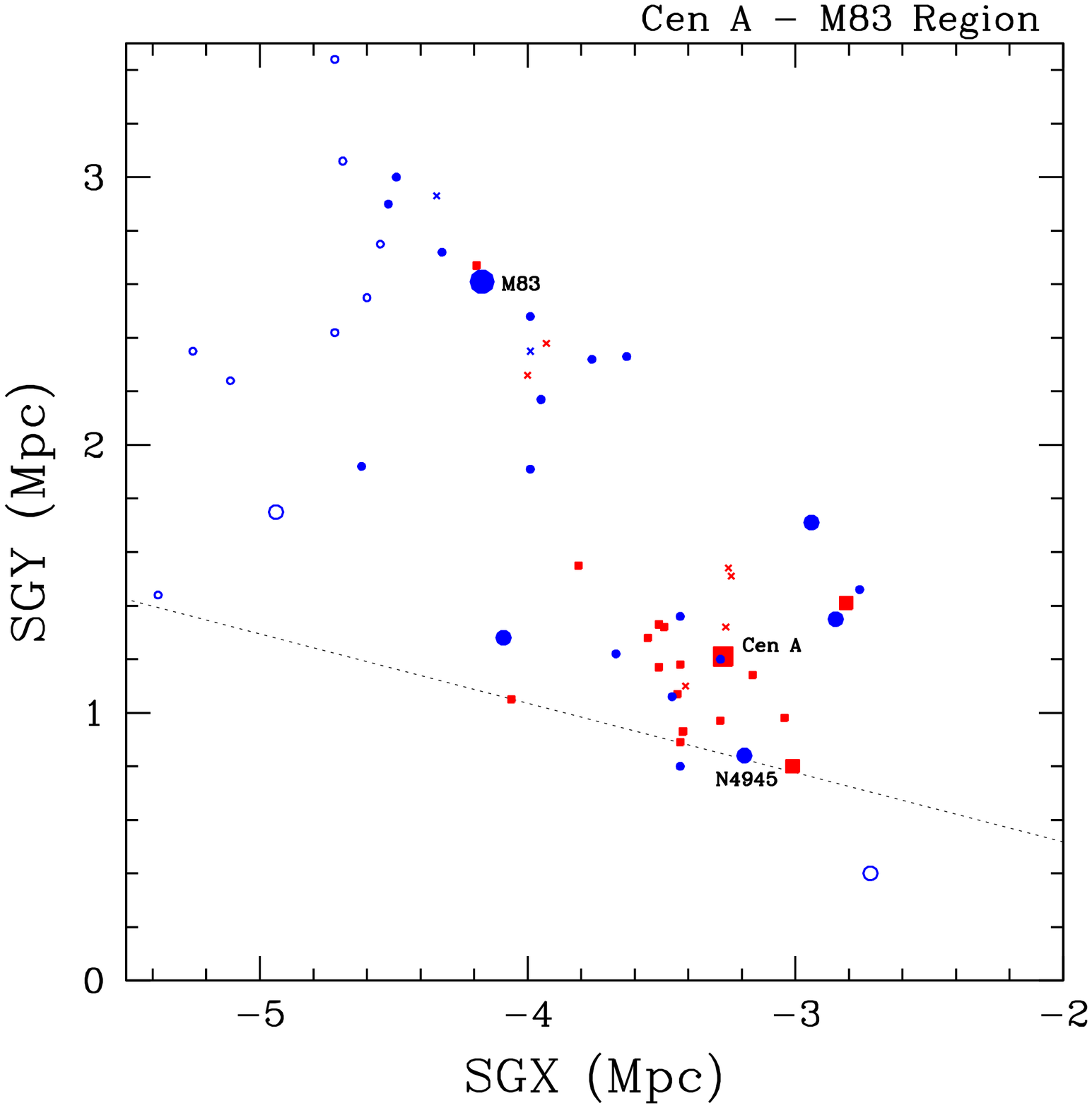}
\includegraphics[scale=.3]{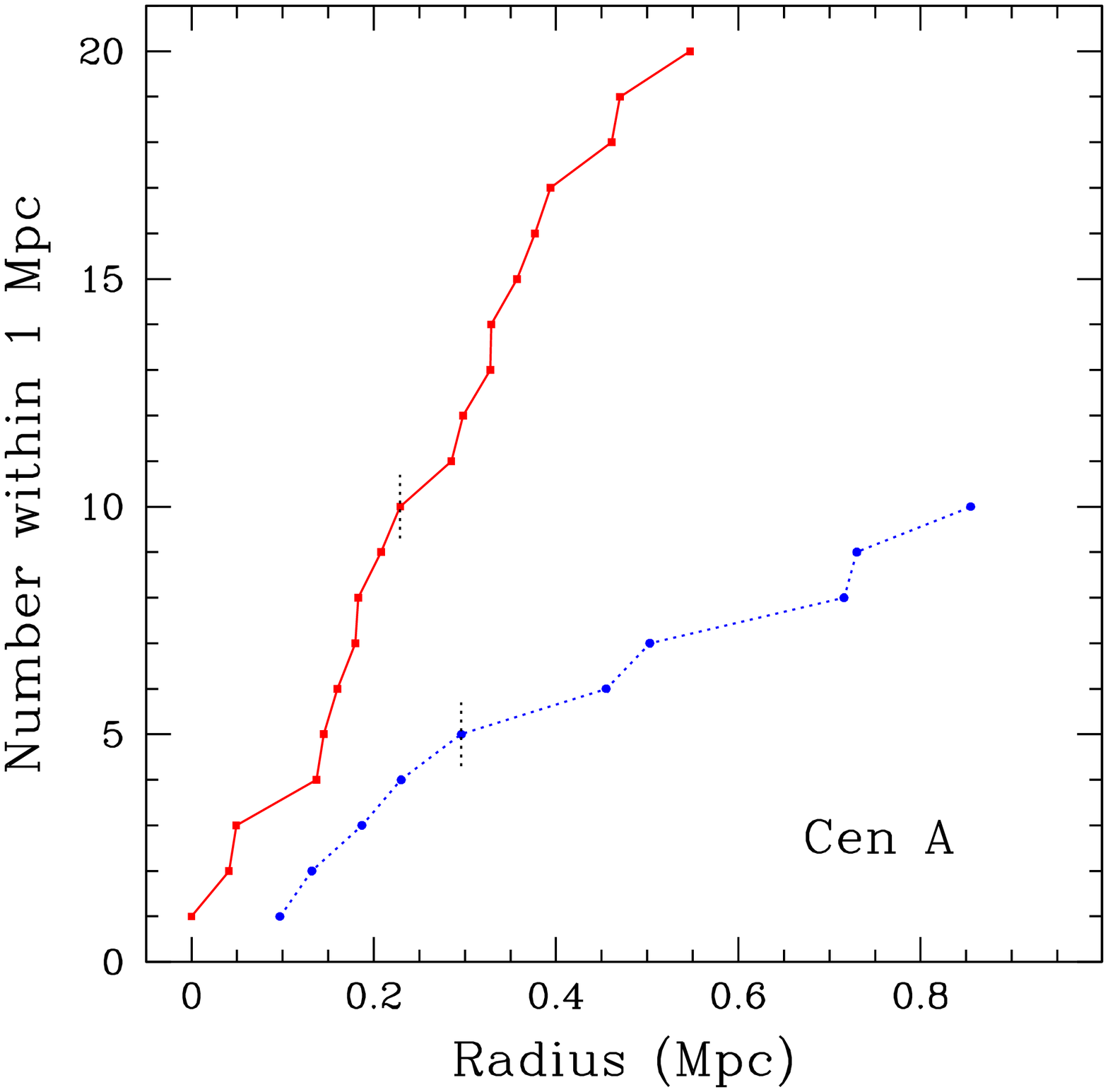}
\caption{{\it Left:} Distribution of galaxies in the region of Cen~A and M83.  The slanted dotted line is a projection of Galactic latitude $+15^{\circ}$.  Accurate distances reveal that Cen A and M83 are the dominant galaxies in two distinct groups.  {\it Right:} The cumulative number of early (solid red) and late (dotted blue) galaxies as a function of projected radius from Cen~A.}
\label{cena} 
\end{figure}

\begin{figure}[]
\sidecaption
\includegraphics[scale=.47]{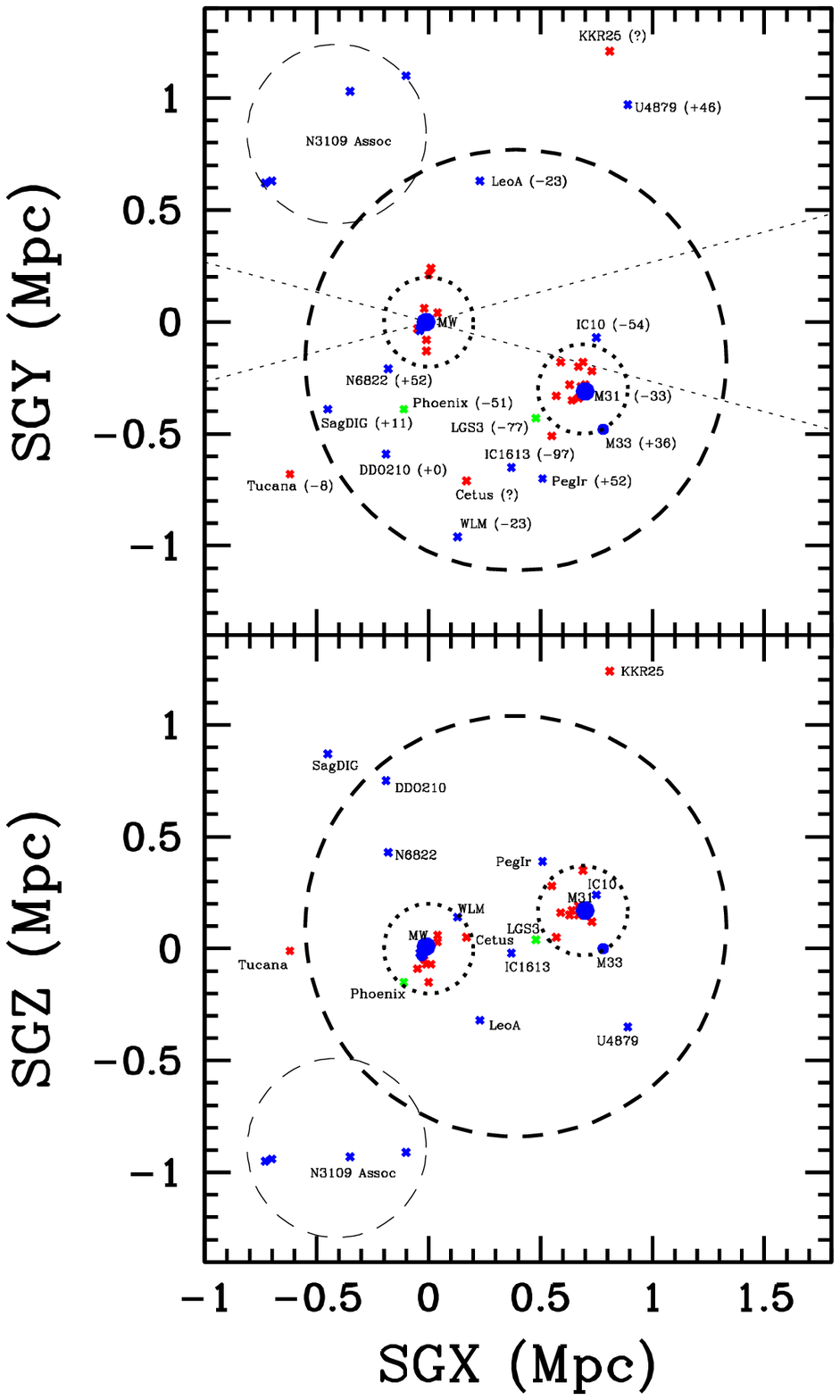}
\includegraphics[scale=.47]{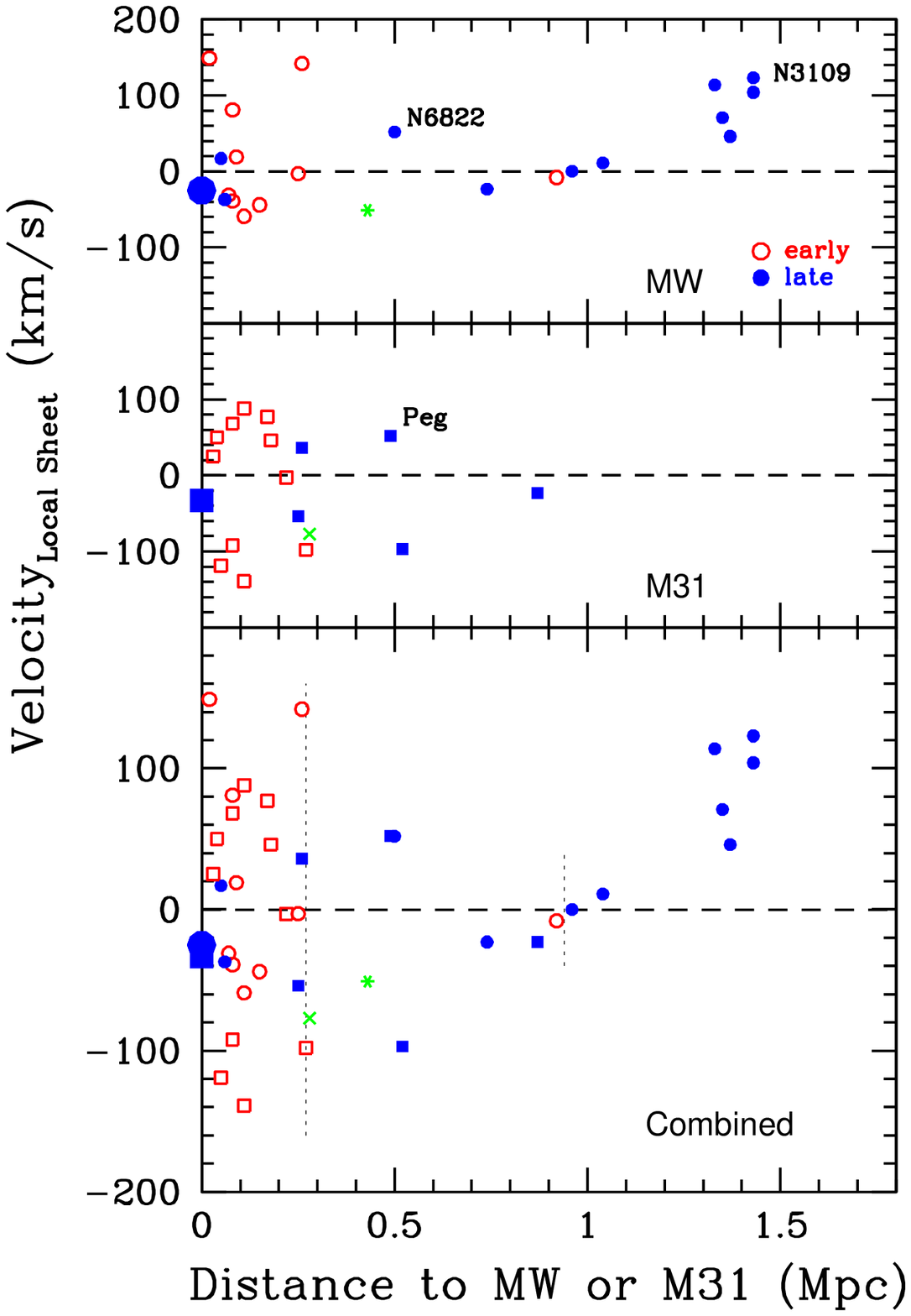}
\caption{{\it Left:} Two projections of the distribution of galaxies around the Local Group.  Blue: late types; red: early types; green: transition.  The slanted dotted lines are projections of Galactic latitude $\pm 15^{\circ}$.  The inner dotted circles approximate the radii of second turnaround for the M31 and Milky Way halos.  The outer heavy dashed circle approximates the first turnaround radius for the combined halos.  A light dashed circle encloses the separate NGC 3109 association. {\it Right:} Velocities as a function of distance from the nearer of M31 or the Milky Way.  Inside the caustic of second turnaround at roughly 270 kpc around each of the major galaxies the velocity dispersion is large and galaxies generally have early types.  Outside second turnaround most galaxies are late types and there is a pattern of infall with the first turnaround located at 940 kpc.}
\label{localgroup} 
\end{figure}

\begin{figure}[]
\sidecaption
\includegraphics[scale=.3]{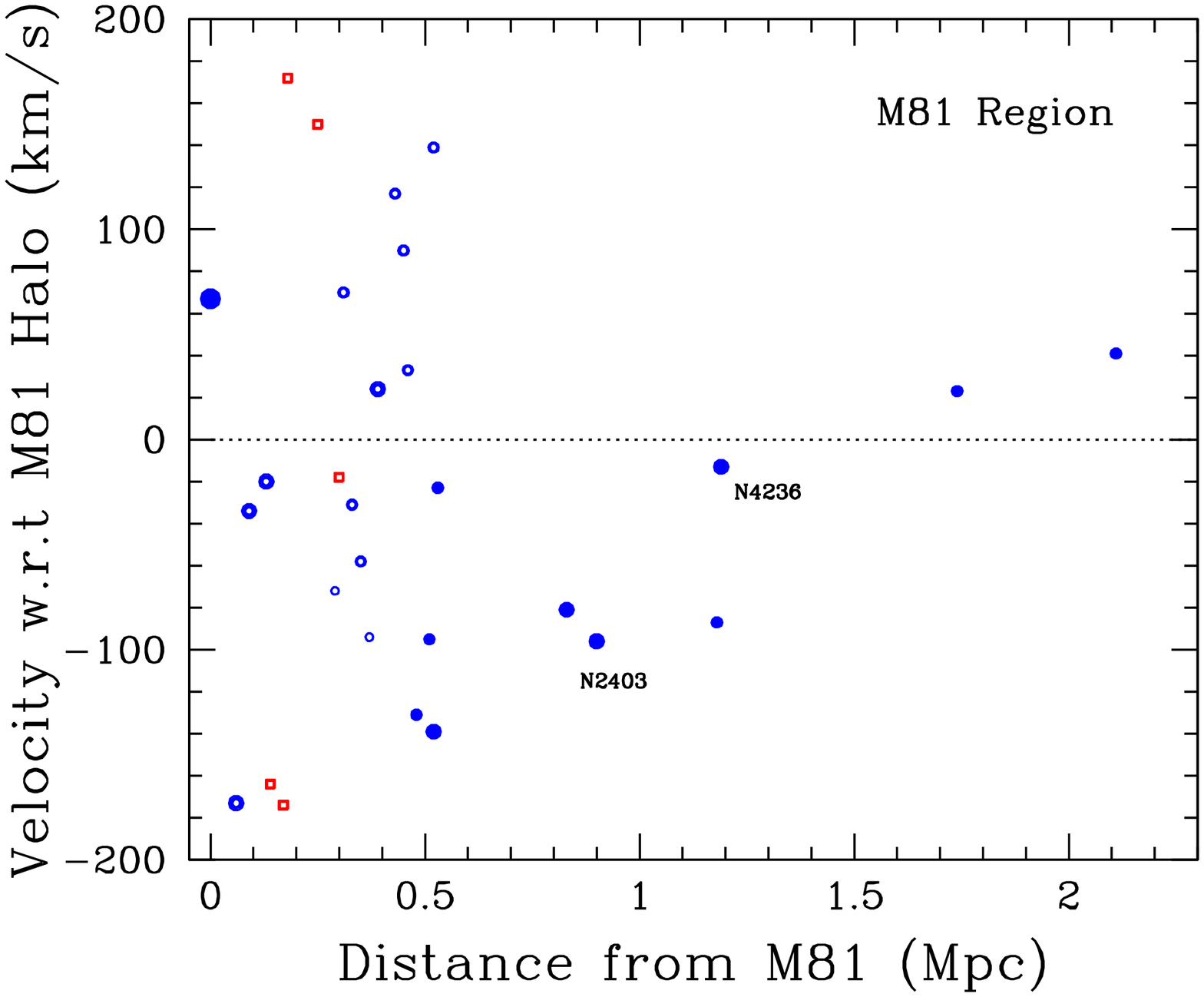}
\includegraphics[scale=.3]{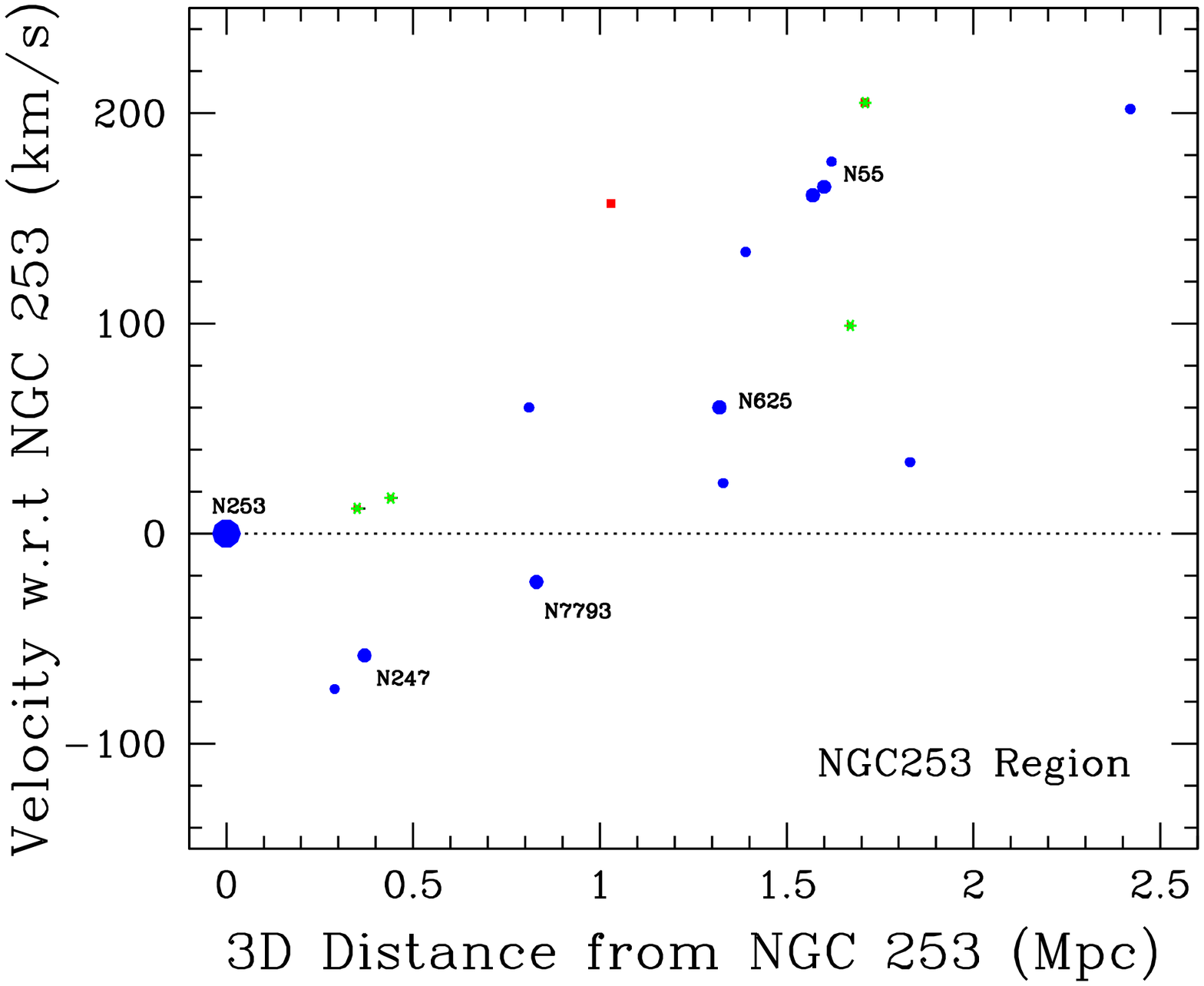}
\caption{Infall patterns onto M81 {\it (Left)} and NGC~253 {\it (Right)}.  With M81 the differentiation between the region of high velocity dispersion within the second turnaround and the region of infall is clear.  The situation is less clear in the case of the smaller halo containing NGC~253.} 
\label{m81-n253} 
\end{figure}

\begin{figure}[]
\sidecaption
\includegraphics[scale=.3]{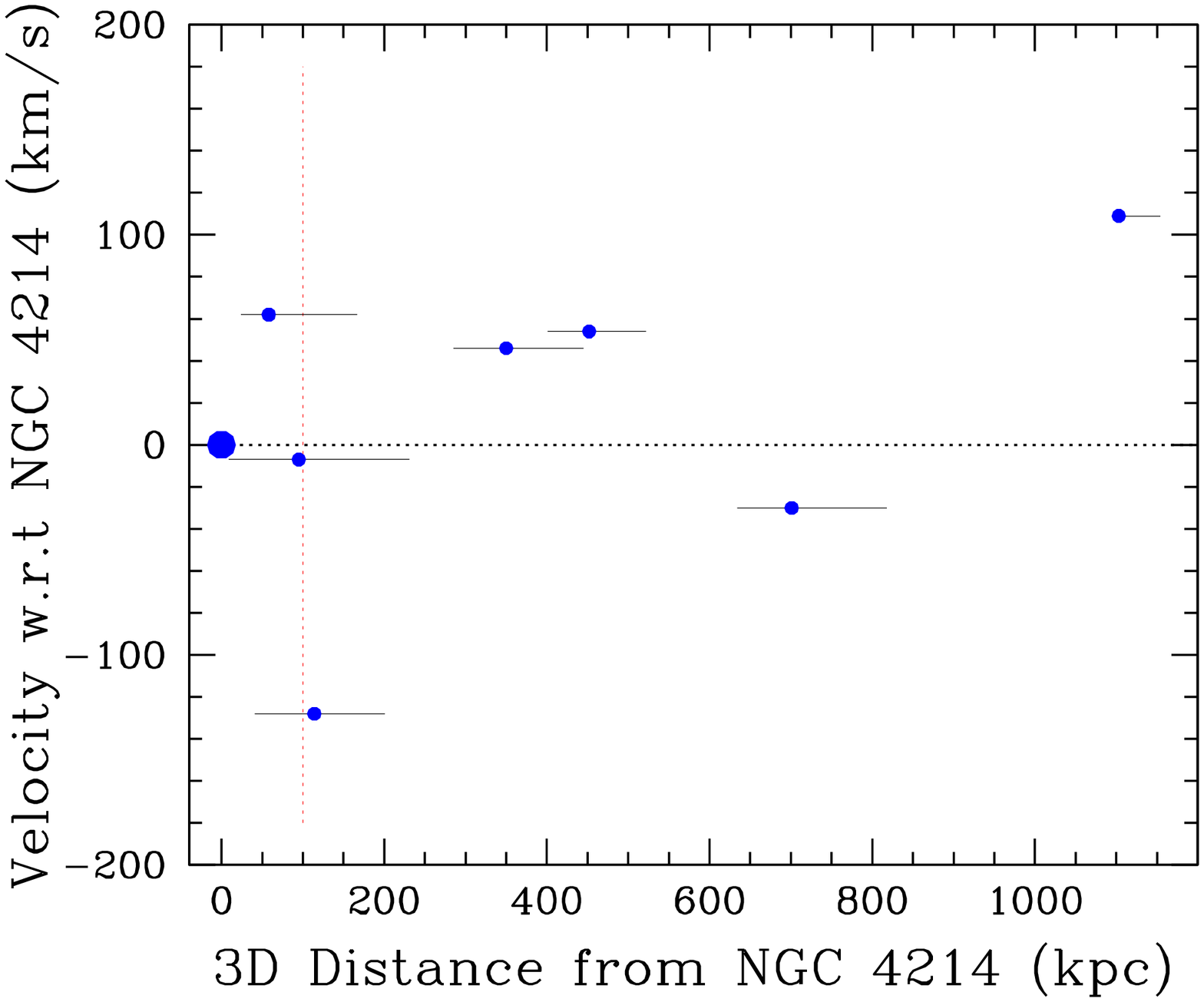}
\includegraphics[scale=.3]{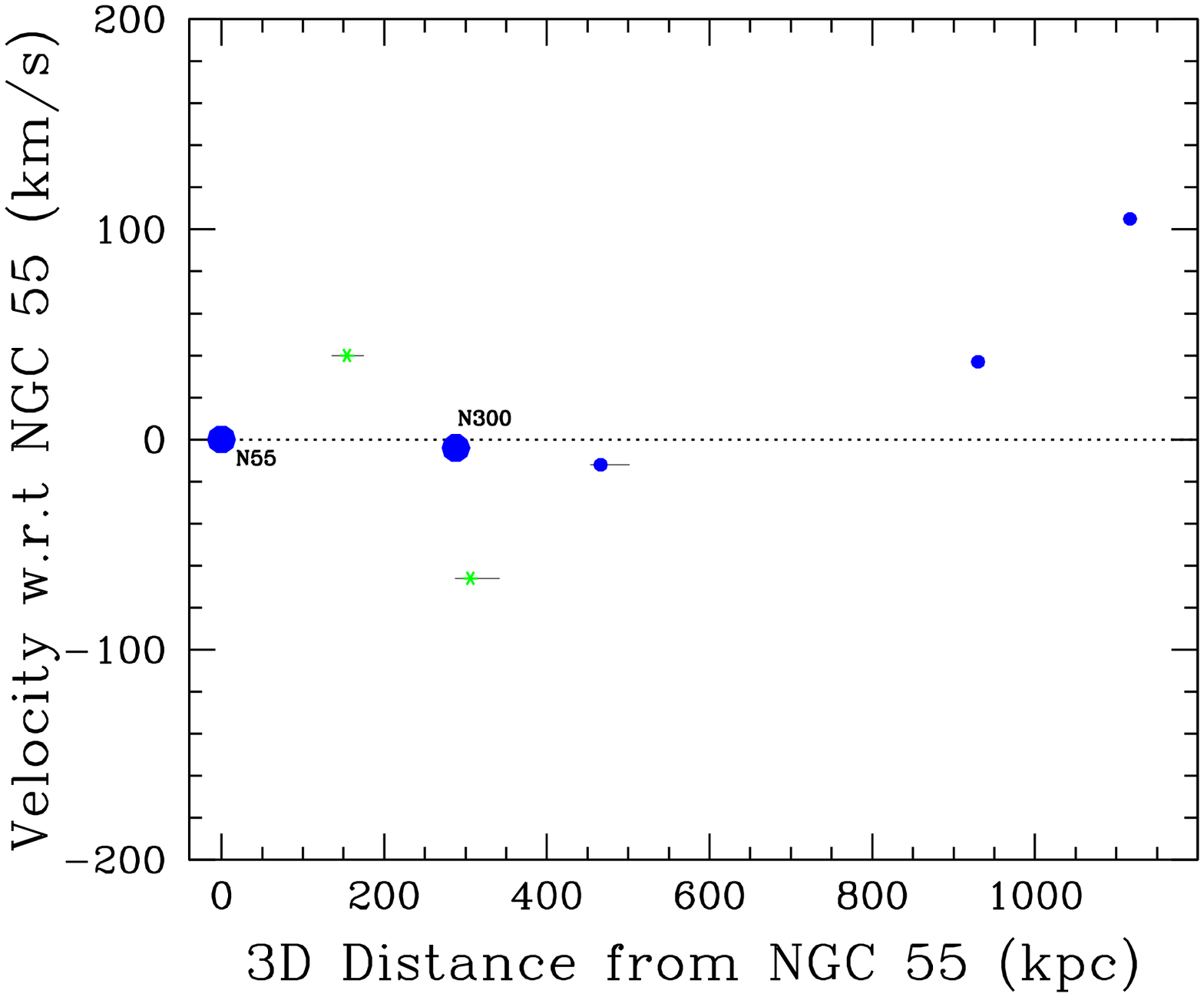}
\caption{Velocities for galaxies in two associations of dwarfs with halo masses below $10^{12}~M_{\odot}$.}
\label{n4214-n55} 
\end{figure}

The very complete information on the three dimensional distribution of galaxies within 4 Mpc lets us look at very small halos.  The largest of the nearby halos is centered on Centaurus~A and has a mass of $1 \cdot 10^{13}~\Msun$.  It is seen in Figure~\ref{cena} that the Cen~A and M83 halos are discrete and the Cen~A halo is host to a substantial family of gas-poor dwarfs \citep{2002A&A...385...21K, 2007AJ....133..504K}.

Figure~\ref{localgroup} gives attention to the historical Local Group, an environment with two discrete halos with masses $1-2 \cdot 10^{12}~\Msun$ within a common infall zone.  There is clear morphological segregation between the regimes inside and outside the second turnaround caustics.  There is  reasonable definition of the surface of first turnaround \citep{2009MNRAS.393.1265K}.

The M81 Group provides a relatively clean situation \citep{2002A&A...383..125K}.  The central halo of $4 \cdot 10^{12}~\Msun$ is characterized by a large dispersion of velocities and it is surrounded by a pattern of infall extending to a first turnaround or zero velocity surface at 1.3~kpc.  See Figure~\ref{m81-n253}.  The second panel of this figure illustrates the more uncertain situation with the $\sim 1 \cdot 10^{12}~\Msun$ halo centered on NGC~253 \citep{1998AJ....116.2873J, 2003A&A...404...93K}.

Finally, Figure~\ref{n4214-n55} illustrates the velocity patterns around two halos in the range $4-7 \cdot 10^{11}~\Msun$ that only contain minor galaxies with luminosities of the LMC or less \citep{2006AJ....132..729T}.  These are interesting because luminosities are so small that mass to light ratios are high.  It is suspected that smaller halos become unidentifiable because of a lack of galaxy signposts.

\section{Scaling Relations}

\begin{figure}[]
\sidecaption
\includegraphics[scale=.3]{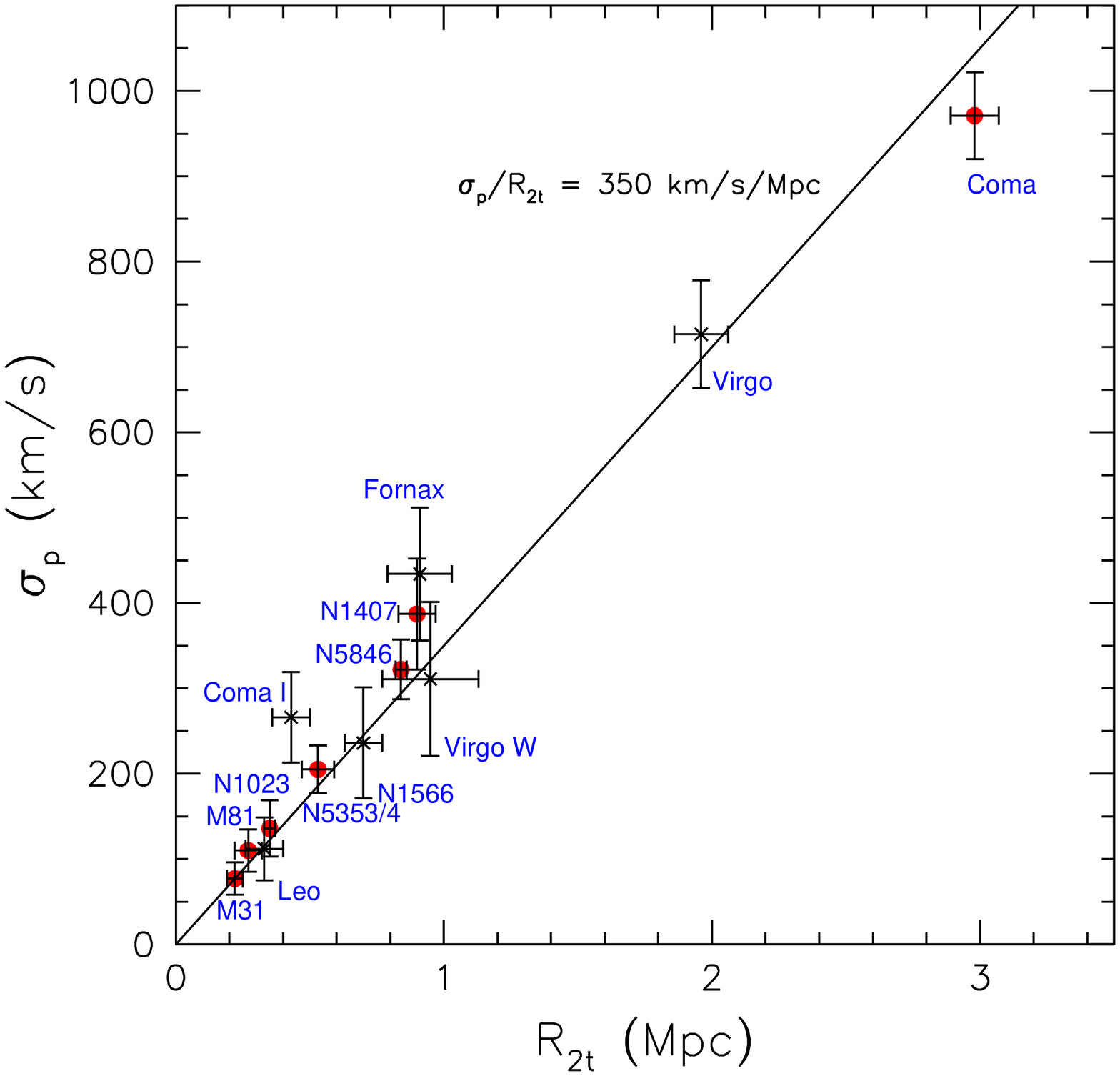}
\includegraphics[scale=.3]{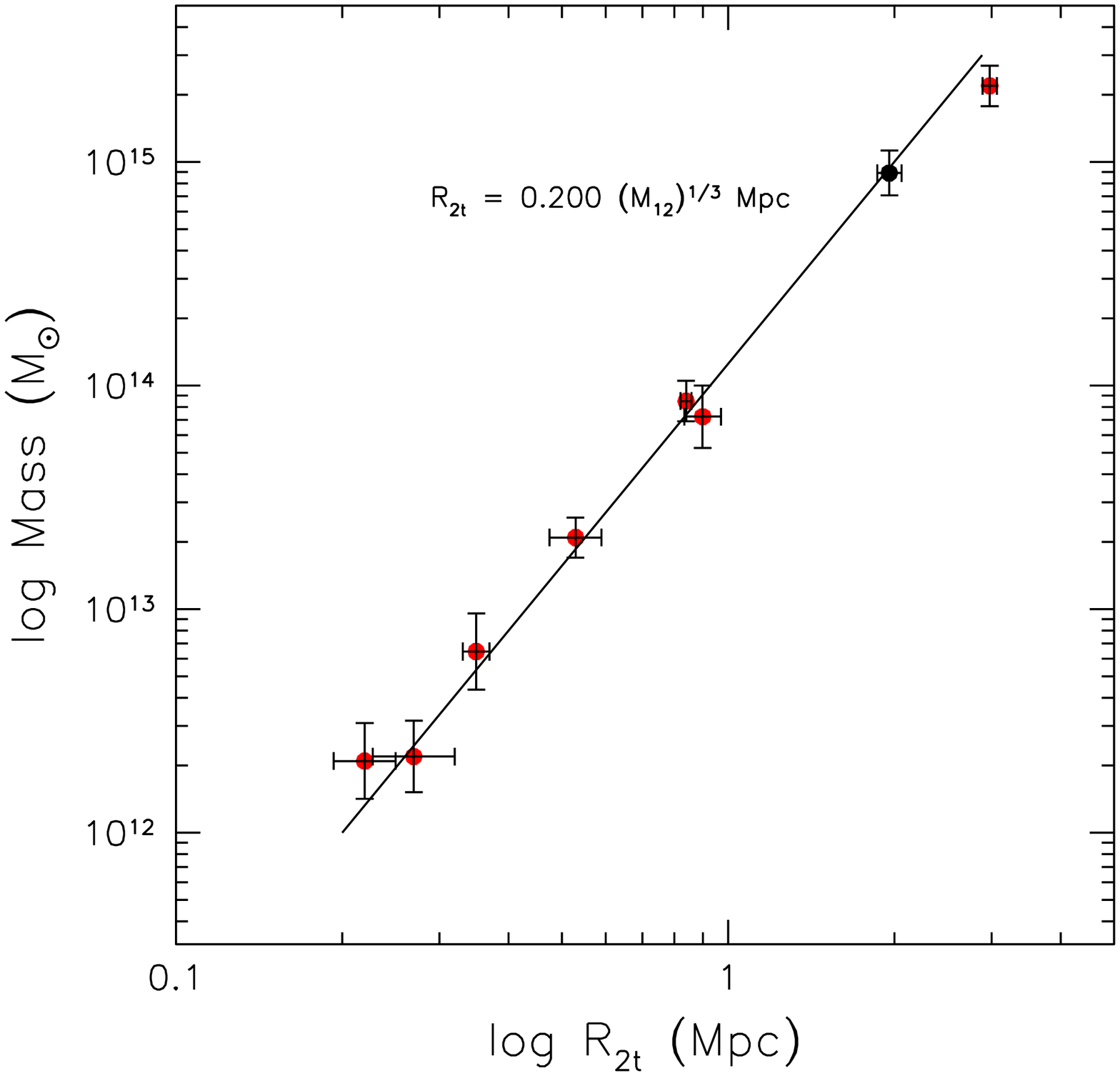}
\caption{{\it Left:} Correlation between projected second turnaround radius and line-of-sight velocity dispersion.  {\it Right:} Correlation between projected second turnaround radius and halo virial mass.}
\label{rv-rm} 
\end{figure}

\begin{figure}[]
\sidecaption
\includegraphics[scale=.3]{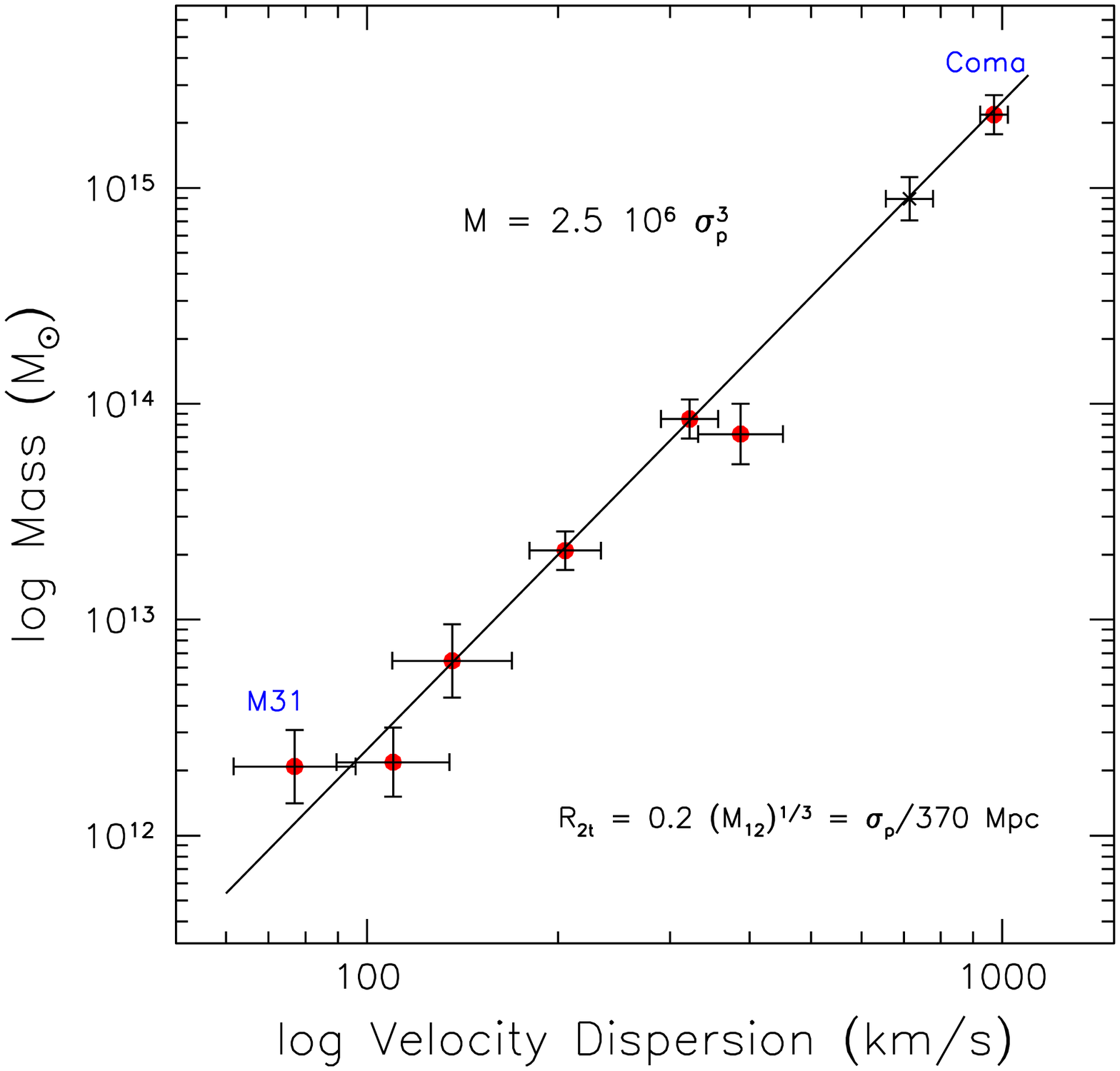}
\includegraphics[scale=.3]{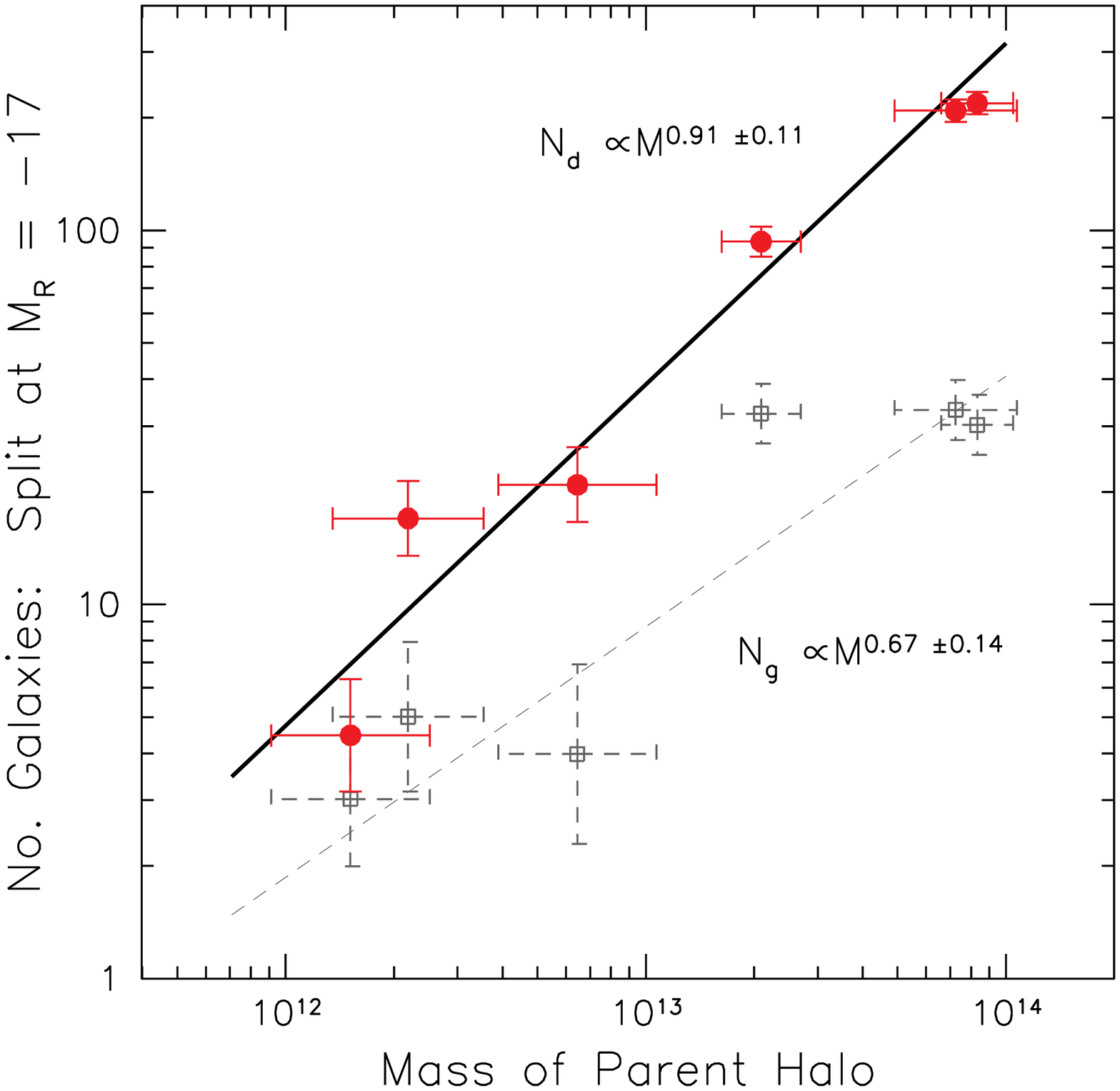}
\caption{{\it Left:} Correlation between the halo virial mass and line-of-sight velocity dispersion. {\it Right:} Number of galaxies as a function of the mass of the parent halo, separated between dwarfs (red points and solid line fit) and giants (grey points and dashed fit).  The separation between giants and dwarfs is made at $M_R = -17$.}
\label{mv-mn} 
\end{figure}

It was anticipated in the introduction that the two independent variables $r_{2t}$ and $\sigma_p$ would be directly correlated.  The relationship is seen in the left panel of Figure~\ref{rv-rm}.  The constant of proportionality is found empirically to be 350 \kmsMpc.

Two interesting relations are shown in the right panel of Figure \ref{rv-rm} and the left panel of Figure~\ref{mv-mn}.  Each involves the virial mass calculated according to
\begin{equation}
M_{\rm v} = {{\sigma_p^2 r_g} \over {G}}
\end{equation}
where the virial radius $r_g$ was defined in Eq.~\ref{rg}
so is calculated independently from $r_{2t}$.  What is being plotted at the right in Fig.~\ref{rv-rm} is $r_{2t}$ versus $\sigma_p^{2/3} r_g^{1/3}$.

The derived correlation is 
\begin{equation}
r_{2t} = 0.200~M_{12}^{1/3}~{\rm Mpc} 
\end{equation}
if the virial mass is in units of $10^{12}~\Msun$.  The tight correlation implies a close correlation between $r_{2t}$ and $r_g$. 
It follows that the correlation seen at the left in Fig.~\ref{mv-mn} would exist.  In this plot the slope is fixed and the zero-point is determined by the relations seen in the two panels of Fig. \ref{rv-rm}.  The mass--dispersion relation for groups obeys the law
\begin{equation}
M_{\rm v} = 2.5 \times 10^6~\sigma_p^3
\end{equation}
where mass $M_{\rm v}$ is in solar units and velocity dispersion $\sigma_p$ is in \kms.

\section{The Radius of First Turnaround}

The best opportunities to measure $r_{1t}$, the radius of first turnaround, are presented by the groups within 4 Mpc including the Local Group.  An initial discussion by    \citet{1986ApJ...307....1S} was carried forward most recently by  \citet{2009MNRAS.393.1265K}.  The situation in the Local Group is shown in Figure~\ref{localgroup}.  The value of $r_{1t} = 940$ kpc enclosing a mass of $2 \times 10^{12}~\Msun$ seems reasonably constrained to an uncertainty of 10\%.  However spherical symmetry is a poor approximation for the zero velocity surface of the Local Group because of the dumbbell distribution of mass between M31 and the Milky Way.  Moreover, only 5 galaxies are well placed to define the zero velocity surface.

A second relatively clean opportunity to measure the surface of first turnaround is presented by the pattern of infall around M81 seen in Fig.~\ref{m81-n253}.   The zero velocity surface lies at 1.3 Mpc from the center of a $4 \times 10^{12}~M_{\odot}$ halo.

This limited information nonetheless gives us a preliminary calibration of the relationship between $r_{1t}$ and the mass within this radius:
\begin{equation}
r_{1t} = 0.78~M_{12}^{1/3}~{\rm Mpc}.
\end{equation}

An interesting potential confirmation of the validity of this relation could come from the definition of the zero velocity surface around the Virgo Cluster.  It is anticipated to lie at $r_{1t} \sim 7$ Mpc, a value in accordance with long-standing models of Virgo infall   \citep{1984ApJ...281...31T}.  A dense grid of galaxy distances across the Local Supercluster is needed to resolve the issue.

The relationship between a specific radius like $r_{2t}$ or $r_{1t}$ and the mass internal to the specific radius is virtually independent of the dark energy content of the universe.  However, because of the relative differences with respect to the age of the universe between the onset of collapse that is associated with $r_{2t}$ and the current age associated with $r_{1t}$, the ratio $r_{1t}/r_{2t}$ has a dependence on the dark energy content.

Currently the best estimate of the ratio of these radii is based on observations of the M81 halo.  From the data used to generate the left panel of Fig.~\ref{m81-n253}  a tentative ratio $r_{1t}/r_{2t} = 3.5 \pm 0.4$ is found where the error estimate includes random but not systematic effects.  At present the errors encompass the full range of reasonable cosmological models.

Another quantity of potential interest is the surface of zero gravity, the surface around an overdense region that is at the limit of what is destined to ever collapse onto the overdensity.  This surface is well specified as a function of cosmological model in the case of spherical collapse.  With $\Omega_{\Lambda} = 0.7$ in a flat universe then $r_{ZG} \sim 1.4 r_{1t}$   \citep{2008A&A...488..845P}.  However given the close proximity of separate collapsed halos, often with more than one halo within an infall region and separate infall regions close to overlapping, the assumption of spherical collapse is generally a bad approximation.  It must be a very bad approximation with $r_{ZG}$.

\section{Parent Halos and their Children}

Groups of both the early and late types were studied during the CFHT Megacam imaging campaign.  The completeness limit for membership studies was typically $M_R = -11$ or fainter.  In an effort to study the properties and possible variations in the luminosity function, galaxies were separated into two bins: `giants' with $M_R < -17$ and `dwarfs' with $-17 < M_R < -11$.  Small but significant variations in luminosity functions were found.  The variations could be characterized by variations in the ratio of dwarfs to giants.  The groups of predominantly early types have larger dwarf/giant ratios.  An interesting result is found if the number of giants or dwarfs is plotted against the group virial mass.  The correlation is poor with giants but very pronounced with dwarfs.  The result is shown in Figure~\ref{mv-mn}.  The straight line describes the relation
\begin{equation}
N_d = 5.2~M_{12}^{0.91 \pm 0.11}
\end{equation}
where $N_d$ is the number of dwarfs with $-17 < -M_R <-11$ in the group.  The slope is consistent within the errors with unity.  The number of dwarfs per unit halo mass is roughly constant.  A count of the number of dwarfs is a good measure of the mass of a group halo.

It is evident from Figs.~\ref{n1023}, \ref{cena}, and \ref{localgroup} that the morphological separation with environment familiar from studies of rich clusters is also seen in small groups.  The caustic of second turnaround is a clear marker of the transition between preponderant morphological types.
Secondary galaxies must loose their supplies of cold gas within of order the halo crossing time of 
\begin{equation}
t_X = {(\pi / 2) R_{2t} \over \sqrt{3} \sigma_p} = 2.5 \times 10^9~{\rm years.} 
\end{equation}

\section{Summary}

Halos can be observationally delineated by a density drop and a transition between dispersed orbits and infall at the radii of second turnaround, $r_{2t}$.  The scaling relations that are theoretically anticipated to exist over a wide range of halo masses are  confirmed from observations of nearby collapsed regions.  Specifically:

$\bullet$ Over the mass range from the halo of M31 to the Coma Cluster there is a correlation between projected second turnaround radius $R_{2t}$ and line-of-sight velocity dispersion $\sigma_p$:  $\sigma_p / R_{2t} = 350$~\kmsMpc.

$\bullet$ Over the same mass range, there is the correlation between $R_{2t}$ and the virial mass $M_{12}$ measured in units of $10^{12}~M_{\odot}$:  $R_{2t} = 0.200 (M_{12})^{1/3}$ Mpc.

$\bullet$ These two correlations can be combined to give the alternative, but not independent, relation:   $M_{\rm v} / M_{\odot} = 2.5 \times 10^6 \sigma_p^3$.

Dwarf galaxies serve to identify collapsed halos because they are numerous.  Dwarfs and giants alike that are located within the radii of second turnaround tend to be gas-poor `early' types while galaxies of all sizes outside the second turnaround tend to be gas-rich `late' types.  The transition between predominantly early and late populations is another indicator of the radius of second turnaround.

Dwarf galaxies delineate collapsed halos in a way that is surprisingly quantitative.  If the number of dwarf galaxies with $-17 < M_R < -11$ are counted, then the number of these dwarfs in a halo depends linearly on the mass of the halo.  Count the number of dwarfs and one has a measure of the mass of the halo.

There are differences in the luminosity function of galaxies with environment \citep{2009MNRAS.398..722T}.  More dynamically evolved environments have somewhat steeper faint end slopes.  However, it appears that the reason is not more dwarfs per unit mass in evolved environments but fewer intermediate luminosity systems in the vicinity of $L^{\star}$, the luminosity that characterizes the exponential cutoff from the faint end power law distribution.  It is suspected that intermediate luminosity systems are being lost through mergers with the central dominant galaxy.  At faint luminosities, the production and depletion mechanisms are such that the number of dwarfs per unit halo mass remains roughly constant.

\begin{acknowledgement}
My collaborators in this enterprise are Kristin Chiboucas, H\'el\`ene Courtois, Brad Jacobs, Igor Karachentsev, Andisheh Mahdavi, Luca Rizzi, Ed Shaya, and Neil Trentham.  A major wide field imaging program has been carried out with the Canada-France-Hawaii Telescope.  Spectroscopic follow up has been carried out with the Subaru and Keck telescopes.  Another major part of the program has involved observations with Hubble Space Telescope.  Support has been provided by the Space Telescope Science Institute in connection with the satellite observations and by the National Science Foundation through the grants AST0307706 and NSF0908846. 
\end{acknowledgement}
%

%\input{referenc}

%\begin{thebibliography}{99.}
% Use the following syntax and markup for your references if 
% the subject of your book is from the field 
% "Mathematics, Physics, Statistics, Computer Science"
%

% Journal article
%\bibitem{phys-journal} S. Preuss, A. Demchuk Jr., M. Stuke, Appl. Phys. A \textbf{61}
%
%\end{thebibliography}

\bibliographystyle{apj}
\bibliography{tully}

\end{document}

%% file: defs.tex
\newcommand{\kms}{km~s$^{-1}$}
\newcommand{\Msun}{M_{\odot}}
\newcommand{\Lsun}{L_{\odot}}
\newcommand{\ML}{M_{\odot}/L_{\odot}}
\newcommand{\etal}{{et al.}\ }
\newcommand{\hhh}{h_{100}}
\newcommand{\hsq}{h_{100}^{-2}}
\newcommand{\tn}{\tablenotemark}
\newcommand{\mdot}{\dot{M}}
\newcommand{\p}{^\prime}
\newcommand{\kmsMpc}{km~s$^{-1}$~Mpc$^{-1}$}

%% file: tully.bbl
\begin{thebibliography}{27}
\expandafter\ifx\csname natexlab\endcsname\relax\def\natexlab#1{#1}\fi

\bibitem[{{Bertschinger}(1985)}]{1985ApJS...58...39B}
{Bertschinger}, E. 1985, ApJS, 58, 39

\bibitem[{{Binggeli} {et~al.}(1987){Binggeli}, {Tammann}, \&
  {Sandage}}]{1987AJ.....94..251B}
{Binggeli}, B., {Tammann}, G.~A., \& {Sandage}, A. 1987, AJ, 94, 251

\bibitem[{{Boylan-Kolchin} {et~al.}(2009){Boylan-Kolchin}, {Springel}, {White},
  {Jenkins}, \& {Lemson}}]{2009MNRAS.398.1150B}
{Boylan-Kolchin}, M., {Springel}, V., {White}, S.~D.~M., {Jenkins}, A., \&
  {Lemson}, G. 2009, MNRAS, 398, 1150

\bibitem[{{Chernin} {et~al.}(2009){Chernin}, {Teerikorpi}, {Valtonen},
  {Dolgachev}, {Domozhilova}, \& {Byrd}}]{2009A&A...507.1271C}
{Chernin}, A.~D., {Teerikorpi}, P., {Valtonen}, M.~J., {Dolgachev}, V.~P.,
  {Domozhilova}, L.~M., \& {Byrd}, G.~G. 2009, A\&A, 507, 1271

\bibitem[{{Diaferio} \& {Geller}(1997)}]{1997ApJ...481..633D}
{Diaferio}, A. \& {Geller}, M.~J. 1997, ApJ, 481, 633

\bibitem[{{Gunn} \& {Gott}(1972)}]{1972ApJ...176....1G}
{Gunn}, J.~E. \& {Gott}, J.~R.~I. 1972, ApJ, 176, 1

\bibitem[{{Jacobs} {et~al.}(2009){Jacobs}, {Rizzi}, {Tully}, {Shaya},
  {Makarov}, \& {Makarova}}]{2009AJ....138..332J}
{Jacobs}, B.~A., {Rizzi}, L., {Tully}, R.~B., {Shaya}, E.~J., {Makarov}, D.~I.,
  \& {Makarova}, L. 2009, AJ, 138, 332

\bibitem[{{Jerjen} {et~al.}(1998){Jerjen}, {Freeman}, \&
  {Binggeli}}]{1998AJ....116.2873J}
{Jerjen}, H., {Freeman}, K.~C., \& {Binggeli}, B. 1998, AJ, 116, 2873

\bibitem[{{Karachentsev} {et~al.}(2002{\natexlab{a}}){Karachentsev}, {Dolphin},
  {Geisler}, {Grebel}, {Guhathakurta}, {Hodge}, {Karachentseva}, {Sarajedini},
  {Seitzer}, \& {Sharina}}]{2002A&A...383..125K}
{Karachentsev}, I.~D., {Dolphin}, A.~E., {Geisler}, D., {Grebel}, E.~K.,
  {Guhathakurta}, P., {Hodge}, P.~W., {Karachentseva}, V.~E., {Sarajedini}, A.,
  {Seitzer}, P., \& {Sharina}, M.~E. 2002{\natexlab{a}}, A\&A, 383, 125

\bibitem[{{Karachentsev} {et~al.}(2003){Karachentsev}, {Grebel}, {Sharina},
  {Dolphin}, {Geisler}, {Guhathakurta}, {Hodge}, {Karachentseva}, {Sarajedini},
  \& {Seitzer}}]{2003A&A...404...93K}
{Karachentsev}, I.~D., {Grebel}, E.~K., {Sharina}, M.~E., {Dolphin}, A.~E.,
  {Geisler}, D., {Guhathakurta}, P., {Hodge}, P.~W., {Karachentseva}, V.~E.,
  {Sarajedini}, A., \& {Seitzer}, P. 2003, A\&A, 404, 93

\bibitem[{{Karachentsev} {et~al.}(2004){Karachentsev}, {Karachentseva},
  {Huchtmeier}, \& {Makarov}}]{2004AJ....127.2031K}
{Karachentsev}, I.~D., {Karachentseva}, V.~E., {Huchtmeier}, W.~K., \&
  {Makarov}, D.~I. 2004, AJ, 127, 2031

\bibitem[{{Karachentsev} {et~al.}(2009){Karachentsev}, {Kashibadze}, {Makarov},
  \& {Tully}}]{2009MNRAS.393.1265K}
{Karachentsev}, I.~D., {Kashibadze}, O.~G., {Makarov}, D.~I., \& {Tully}, R.~B.
  2009, MNRAS, 393, 1265

\bibitem[{{Karachentsev} {et~al.}(2002{\natexlab{b}}){Karachentsev}, {Sharina},
  {Dolphin}, {Grebel}, {Geisler}, {Guhathakurta}, {Hodge}, {Karachentseva},
  {Sarajedini}, \& {Seitzer}}]{2002A&A...385...21K}
{Karachentsev}, I.~D., {Sharina}, M.~E., {Dolphin}, A.~E., {Grebel}, E.~K.,
  {Geisler}, D., {Guhathakurta}, P., {Hodge}, P.~W., {Karachentseva}, V.~E.,
  {Sarajedini}, A., \& {Seitzer}, P. 2002{\natexlab{b}}, A\&A, 385, 21

\bibitem[{{Karachentsev} {et~al.}(2007){Karachentsev}, {Tully}, {Dolphin},
  {Sharina}, {Makarova}, {Makarov}, {Sakai}, {Shaya}, {Kashibadze},
  {Karachentseva}, \& {Rizzi}}]{2007AJ....133..504K}
{Karachentsev}, I.~D., {Tully}, R.~B., {Dolphin}, A., {Sharina}, M.,
  {Makarova}, L., {Makarov}, D., {Sakai}, S., {Shaya}, E.~J., {Kashibadze},
  O.~G., {Karachentseva}, V., \& {Rizzi}, L. 2007, AJ, 133, 504

\bibitem[{{Lynden-Bell}(1967)}]{1967MNRAS.136..101L}
{Lynden-Bell}, D. 1967, MNRAS, 136, 101

\bibitem[{{Mahdavi} {et~al.}(1999){Mahdavi}, {Geller}, {B{\"o}hringer},
  {Kurtz}, \& {Ramella}}]{1999ApJ...518...69M}
{Mahdavi}, A., {Geller}, M.~J., {B{\"o}hringer}, H., {Kurtz}, M.~J., \&
  {Ramella}, M. 1999, ApJ, 518, 69

\bibitem[{{Mahdavi} {et~al.}(2005){Mahdavi}, {Trentham}, \&
  {Tully}}]{2005AJ....130.1502M}
{Mahdavi}, A., {Trentham}, N., \& {Tully}, R.~B. 2005, AJ, 130, 1502

\bibitem[{{Peirani} \& {de Freitas Pacheco}(2008)}]{2008A&A...488..845P}
{Peirani}, S. \& {de Freitas Pacheco}, J.~A. 2008, A\&A, 488, 845

\bibitem[{{Rines} {et~al.}(2003){Rines}, {Geller}, {Kurtz}, \&
  {Diaferio}}]{2003AJ....126.2152R}
{Rines}, K., {Geller}, M.~J., {Kurtz}, M.~J., \& {Diaferio}, A. 2003, AJ, 126,
  2152

\bibitem[{{Sandage}(1986)}]{1986ApJ...307....1S}
{Sandage}, A. 1986, ApJ, 307, 1

\bibitem[{{Shandarin} \& {Zeldovich}(1989)}]{1989RvMP...61..185S}
{Shandarin}, S.~F. \& {Zeldovich}, Y.~B. 1989, Reviews of Modern Physics, 61,
  185

\bibitem[{{Trentham} \& {Tully}(2009)}]{2009MNRAS.398..722T}
{Trentham}, N. \& {Tully}, R.~B. 2009, MNRAS, 398, 722

\bibitem[{{Trentham} {et~al.}(2006){Trentham}, {Tully}, \&
  {Mahdavi}}]{2006MNRAS.369.1375T}
{Trentham}, N., {Tully}, R.~B., \& {Mahdavi}, A. 2006, MNRAS, 369, 1375

\bibitem[{{Tully} {et~al.}(2006){Tully}, {Rizzi}, {Dolphin}, {Karachentsev},
  {Karachentseva}, {Makarov}, {Makarova}, {Sakai}, \&
  {Shaya}}]{2006AJ....132..729T}
{Tully}, R.~B., {Rizzi}, L., {Dolphin}, A.~E., {Karachentsev}, I.~D.,
  {Karachentseva}, V.~E., {Makarov}, D.~I., {Makarova}, L., {Sakai}, S., \&
  {Shaya}, E.~J. 2006, AJ, 132, 729

\bibitem[{{Tully} \& {Shaya}(1984)}]{1984ApJ...281...31T}
{Tully}, R.~B. \& {Shaya}, E.~J. 1984, ApJ, 281, 31

\bibitem[{{Tully} \& {Trentham}(2008)}]{2008AJ....135.1488T}
{Tully}, R.~B. \& {Trentham}, N. 2008, AJ, 135, 1488

\bibitem[{{Vogelsberger} {et~al.}(2009){Vogelsberger}, {White}, {Mohayaee}, \&
  {Springel}}]{2009MNRAS.400.2174V}
{Vogelsberger}, M., {White}, S.~D.~M., {Mohayaee}, R., \& {Springel}, V. 2009,
  MNRAS, 400, 2174

\end{thebibliography}
